\documentclass[normal]{sim}
\usepackage{graphicx}
\usepackage{txfonts}
\usepackage{natbib}

\begin{document}

\title{Investigating the potential of the Pan-Planets project using
  Monte Carlo simulations}

\author{J.~Koppenhoefer\inst{1,2}, C.~Afonso\inst{3},
  R.~P.~Saglia\inst{1,2} and Th.~Henning\inst{3}}

\institute{University Observatory Munich, Scheinerstrasse 1, 81679
  M\"unchen, Germany \email{koppenh@usm.uni-muenchen.de} \and Max
  Planck Institute for Extraterrestrial Physics, Giessenbachstrasse,
  85748 Garching, Germany \email{koppenh@mpe.mpg.de,
    saglia@mpe.mpg.de} \and Max Planck Institute for Astronomy,
  K\"onigstuhl 17, 69117 Heidelberg, Germany \email{afonso@mpia.de,
    henning@mpia.de} }

\date{Accepted December 2, 2008}

\abstract{Using Monte Carlo simulations we analyze the potential of
  the upcoming transit survey Pan-Planets. The analysis covers the
  simulation of realistic light curves (including the effects of
  ingress/egress and limb-darkening) with both correlated and
  uncorrelated noise as well as the application of a
  box-fitting-least-squares detection algorithm. In this work we show
  how simulations can be a powerful tool in defining and optimizing
  the survey strategy of a transiting planet survey. We find the
  Pan-Planets project to be competitive with all other existing and
  planned transit surveys with the main power being the large 7 square
  degree field of view. In the first year we expect to find up to 25
  Jupiter-sized planets with periods below 5 days around stars
  brighter than V = 16.5 mag. The survey will also be sensitive to
  planets with longer periods and planets with smaller radii. After
  the second year of the survey, we expect to find up to 9 Warm
  Jupiters with periods between 5 and 9 days and 7 Very Hot Saturns
  around stars brighter than V = 16.5 mag as well as 9 Very Hot
  Neptunes with periods from 1 to 3 days around stars brighter than i'
  = 18.0 mag.}


\titlerunning{Pan-Planets: expected number of detections}

\maketitle
%
%
%
%
\section{Introduction}
Thirteen years after the first discovery of a planet revolving around
a main-sequence star \citep{1995Natur.378..355M} more than 300 extra
solar planets are known. The majority of these have been detected by
looking for periodic small amplitude variations in the radial velocity
of the planet's host star. This method reveals the period, semi-major
axis and eccentricity of the planetary orbit, but due to a generally
unknown inclination only the minimum mass of the planet can be
derived. The situation is different in the case of a planet that
transits its host star. The planet blocks a small fraction of the
stellar surface resulting in a periodic drop in brightness. In
combination with radial velocity measurements the light curve provides
many additional parameters like inclination, true mass and radius of
the planet.\\
Transiting planets are subject of many detailed follow-up studies such
as measurement of thermal emission using the secondary transit
\citep{2008arXiv0802.0845C,2008ApJ...673..526K} or measurement of the
spin-orbit alignment using the Rossiter-McLaughlin effect
\citep[e.g.][]{2007ApJ...665L.167W}.\\
In the past years many transit projects have monitored hundreds of
thousands of stars looking for periodic drops in the light curves. A
total of about 50 transiting planets are known to date. Remarkably,
more than half of the transiting planets have been found in the past
year making the transit method equally successful in
that period compared to the radial velocity method.\\
The majority of the recently detected transiting planets have been
found by wide-angle surveys targeting bright stars such as WASP, HAT,
TrES or XO
\citep{2006PASP..118.1407P,2008ApJ...673L..79N,2007ApJ...663L..37O,2005PASP..117..783M}.
Also the space mission Corot has contributed by adding four new
discoveries \citep{2008ASPC..384..270A}. Deep surveys like OGLE
targeting highly crowded regions of the Milky Way disk have not been
able to keep up with the increased detection rate of all-sky
monitoring programs mainly due to limited amount of observation
time and a lower number of target stars.\\
In 2009 Pan-Planets - a new deep transit survey - will start taking
first observations. This project will be more powerful than all
existing deep surveys because of its by far larger field of view,
bigger telescope and faster readout.\\\\
With the first detections of transiting extra-solar planets, several
groups have started to predict the number of planets that could be
found by existing and planned surveys. First estimates were based on
optimistic assumptions and have been mostly over-predictions
\citep[e.g.][]{2003ASPC..294..361H}. For example the frequency of very
close-in planets had been extrapolated from the metallicity biased
results of the radial velocity surveys. Late type dwarfs with higher
metallicity turned out to have a higher frequencies of close-in
planets \citep{2005ApJ...622.1102F} and therefore transit surveys find
less planets compared to radial velocity surveys due to a lower
average metallicity \citep{2006AcA....56....1G}. It was further
assumed that planets could be found around all stars in the target
fields whereas planets transiting giants show a much too faint
photometric signal due to the larger radius of the star. In addition,
the efficiency of the detection algorithm was not taken into account
and all light curves with 2 visible transits were assumed to lead to
a detection which is not the case.\\
More realistic methods have been introduced by
\citet{2003AcA....53..213P} and \citet{2006AcA....56....1G}. Both
groups use an analytical approach assuming a stellar and a planetary
distribution and integrating over period, stellar mass, planetary
radius and volume probed taking into account the detection
probability. Similarly, \citet{2007A&A...475..729F} modeled the OGLE
survey and compared the predicted distributions to the parameters
actually found by OGLE. In a recent study, \citet{2008arXiv0804.1150B}
generalized the formalism of \citet{2006AcA....56....1G} in order to
provide a method that can be used to calculate planet yields for any
photometric survey given the survey parameters like number of nights
observed, bandpass, exposure time, telescope aperture, etc. They
applied their method to a number of different planned surveys like
SDSS-II and the Pan-STARSS 3$\pi$ survey.\\
In this work we use Monte Carlo simulations to predict the number of
planets of the Pan-Planets survey. Our approach is quite general and
applicable to any transit survey. Based on stellar and planetary
populations we model the survey by constructing realistic light curves
and running a detection algorithm on them. In this way we are able to
directly include the effects of limb darkening, ingress/egress and
observational window functions which have not been included in most
previous studies. In addition we introduce a model for correlated
noise and study its impact on the efficiency of the detection
algorithm.\\
To optimize the survey strategy of Pan-Planets we want to address the
following questions: What is the best observing block size (1h or 3h)
and how many fields (3 to 7) should we observe? Given the optimized
survey strategy, we study how many Very Hot Juptiters (VHJ) and Hot
Jupiters (HJ) are expected in the first year and what is the potential
of Pan-Planets to find planets with longer periods, such as Warm
Jupiters (WJ) or planets with smaller radii, such as Very Hot Saturns
(VHS) and Very Hot Neptunes (VHN). We further study whether it will be
more efficient to observe the same target fields in the second year of 
the Pan-Planets survey or to choose new ones.\\\\
In \S\ref{overview} we give a brief overview of the Pan-Planets
survey. \S\ref{simulations} describes in detail the simulations we
performed.  We present our results in \S\ref{results}.  In order to
verify our results we perform a consistency check with the OGLE-III
survey by comparing our predicted yield with the actual number of
planets found (\S\ref{consistency}). Finally we draw our conclusions
in \S\ref{conclusions}. 
\section{Pan-Planets overview}
\label{overview}
The Panoramic Survey Telescope and Rapid Response System (PanSTARRS)
is an Air Force funded project aiming at the detection of killer
asteroids that have the potential of hitting the Earth in the near
future. The prototype mission PanSTARRS1 is using a 1.8m telescope at
the Haleakala Observatories (Maui, Hawaii) to monitor 3$\pi$ of the
sky over a 3.5 yr period starting in early 2009. The telescope is
equipped with the largest CCD camera in the world to date that samples
a field of 7 sq.deg. on a 1.4 Gigapixel array
\citep{2004SPIE.5489...11K} with a pixel-size of 0.258 arcsec.\\
To make use of the large amount of data that will be collected, a
science consortium of institutes from USA, Germany, UK and Taiwan has
defined 12 Key Science Projects, out of which one is the Pan-Planets
transit survey. A total of 120h per year have been dedicated to this
project during the 3.5 yr lifetime of the survey. The actual observing
time will be less due to bad weather and technical downtime. We
account for a 33\% loss in our simulations.\\
In the first 2 years, Pan-Planets will observe 3 to 7 fields in the
direction of the Galactic plane. Exposure and read-out time will be
30s and 10s respectively. The observations will be scheduled in 1h or
3h blocks. The target magnitude range will be 13.5 to 16.5 mag in the
Johnson V-band. The magnitude range is extended to i' = 18 when
searching for Very Hot Neptunes (see \S\ref{subsec.VHN}). More
detailed informations about Pan-Planets are presented in Afonso et
al. (in prep.).
\section{Description of the simulations}
\label{simulations}
The goal of this work is to study the expected number of planets that
will be detected by the Pan-Planets project as a function of different
survey strategies, with a variety of different parameters like number
of fields (3 to 7), length of a single observing block (1h and 3h) and
level of residual red noise (0 mmag, 1 mmag, 2 mmag, 3 mmag and 4
mmag). In total we simulate about 100 different combinations of these
parameters for each of 5 different planet populations (see
\S\ref{subsec.planets}).\\
In our simulations we follow a full Monte-Carlo approach, starting
with the simulation of light curves with realistic transit signals.
Systematic effects coming from data reduction steps on image basis,
such as differential imaging or PSF-photometry are taken into account
by adding non-Gaussian correlated noise, the so called red noise
\citep{2006MNRAS.373..231P}, to our light curves (see
\S\ref{subsec.red}). We apply a box-fitting-least-squares algorithm to
all simulated light curves in order to test whether a transiting
planet is detected or not.\\
For each star in the input stellar distribution (\S\ref{subsec.stars})
we decide randomly whether it has a planet or not, depending on the
fraction of stars having a planet of this type. In the case it has a
planet, we randomly pick a planet from the input planet distribution
(\S\ref{subsec.planets}) and create a star-planet pair which is
attributed a randomly oriented inclination vector resulting in a
transiting or non-transiting orbit (the geometric probability for a
transiting orbit depends on stellar radius and semi-major axis of the
orbit).\\
In the case of a transiting orbit, the light curve is simulated based
on stellar and planetary parameters and the observational dates we
specified (see \S\ref{subsec.obsdates}). The shape of the transit is
calculated according to the formulae of \citet{2002ApJ...580L.171M}
and includes the effects of ingress/egress and limb-darkening. We add
uncorrelated Gaussian (white) and correlated non-Gaussian (red) noise
to our light curves. Details about our noise model are given in
\S\ref{subsec.white} and \S\ref{subsec.red}. After the simulation of
the light curves, we apply our detection algorithm and our detection
cuts as described in \S\ref{subsec.analysis}, and count how many 
planets we detect.\\
One simulation run is finished after each star has been picked
once. In this way one run represents one possible outcome of the
Pan-Planets survey. Since in the majority of cases the star has no
planet or the inclination is such that no transits are visible, there
are in general only a few transiting light curves per run. For each
planet population and each set of survey parameters we simulate
25\,000 runs. For the selected survey strategy we increase the
precision to 100\,000 runs. The numbers we list in our results are
averages over these runs. The scatter of the individual outcomes
allows us to derive errors for our estimates.
\subsection{Input stellar distribution}
\label{subsec.stars}
\begin{figure}
  \centering
  \includegraphics[width=0.4\textwidth]{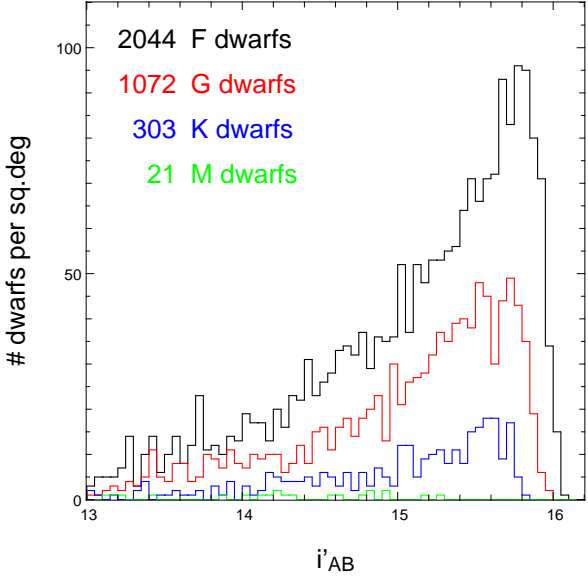}
	\caption{Total number and magnitude histogram for F, G, K and M
    dwarfs (top to bottom) with $m_{i'} \ge$ 13 mag and $m_{V} \le$
    16.5 mag in our target population. Note that the cut on the visual
    magnitude results in an brighter cut in i' for the later type
    stars due to their redder color.}
  \label{mag_hist}
\end{figure}

We make use of a Besan\c{c}on
model\footnote{http://bison.obs-besancon.fr/modele/}
\citep{2003A&A...409..523R} for the spectral type and brightness
distributions of stars in our target fields. A model of 1 sq.deg
centered around RA = $19^h47^m41^s_{\cdot}7$, DEC = $+17^d01^m52^s$ (l
= 54.5, b = -4.2) is scaled to the actual survey area assuming a
constant density. The parameters taken from the model are stellar mass
$M_{star}$, effective temperature $T_{eff}$, surface gravity \mbox{log
  $g$}, metallicity $[Fe/H]$ and apparent
MegaCam\footnote{http://www.cfht.hawaii.edu/Instruments/Imaging/Megacam/}
i'-band AB-magnitude $m_{i'}$. The model also provides colors which we
use to determine the apparent Johnson V-band magnitude $m_{V}$,
according to the following formula derived by
\citet{2002AJ....123.2121S} :
\begin{equation}
  V = g' - 0.55 \cdot (g'-r') - 0.03
\end{equation}
The stellar radii $R_{star}$ are calculated using \mbox{log $g$} and
$M_{star}$ according to $R_{star}$ = sqrt($G \times M_{star}$ /
$g$). Furthermore, $T_{eff}$, \mbox{log $g$}, and $[Fe/H]$ are used to
determine quadratic limb-darkening coefficients according to
\citet{2004A&A...428.1001C} which are based on synthetic ATLAS spectra
\citep{2000A&A...363.1081C}.\\
In total we find 3\,440 F, G, K and M dwarfs\footnote{we refer to
  dwarfs as stars of luminosity class IV-VI} per sq.deg. that are not
saturated (i.e. $m_{i'} \ge$ 13 mag) and are brighter than our radial
velocity follow-up limit (i.e. $m_{V} \le$ 16.5 mag).
\mbox{Fig. \ref{mag_hist}} shows the input stellar distribution.\\
For VHN we extend the target magnitude range to $m_{i'} \le$
18 mag. We find 34\,000 M dwarfs in this range.
\subsection{Input planet distributions}
\label{subsec.planets}
We test five different planetary populations :
\begin{enumerate}
\item{Very Hot Jupiters (VHJ), with radii of 1.0-1.25 $R_{J}$ and
    periods between 1 and 3 days}
\item{Hot Jupiters (HJ), with radii of 1.0-1.25 $R_{J}$ and periods
    between 3 and 5 days}
\item{Warm Jupiters (WJ), with radii of 1.0-1.25 $R_{J}$ and periods
    between 5 and 10 days}
\item{Very Hot Saturns (VHS), with radii of 0.6-0.8 $R_{J}$ and
    periods between 1 and 3 days}
\item{Very Hot Neptunes (VHN), with radii of 0.3 $R_{J}$ and periods
    between 1 and 3 days}
\end{enumerate}
Within the given ranges the radii and periods are homogeneously
distributed.\\
Our predicted yields depend on the frequency of stars that have a
planet for each of the five population. These frequencies are not
known to a very good precision and not many estimates have been
published so far. \citet{2006AcA....56....1G} performed a detailed
study of the OGLE-III survey and derived frequencies of Very Hot
Jupiters and Hot Jupiters by comparing the number of detected planets
in the OGLE-III survey to the number of stars the survey was sensitive
to. They found at 90\% confidence level \mbox{0.1408 $\cdot$
  ($1^{+1.10}_{-0.54}$)\%} of all late type dwarfs to have a VHJ and
\mbox{0.3125 $\cdot$ ($1^{+1.37}_{-0.59}$)\%} to have an HJ.
\citet{2007A&A...475..729F} published comparable results analyzing
the same survey.\\
For VHJ and HJ we use the frequencies published by
\citet{2006AcA....56....1G}. The frequency of WJ we speculate to be
the same as for HJ which is consistent with the OGLE-III results (see
\S\ref{consistency}). Further we assume the frequencies for VHS and
VHN to be 0.714\% (same as for VHJ) and 5\% respectively.
\subsection{White noise model}
\label{subsec.white}
\begin{figure}
  \centering
  \includegraphics[width=0.4\textwidth]{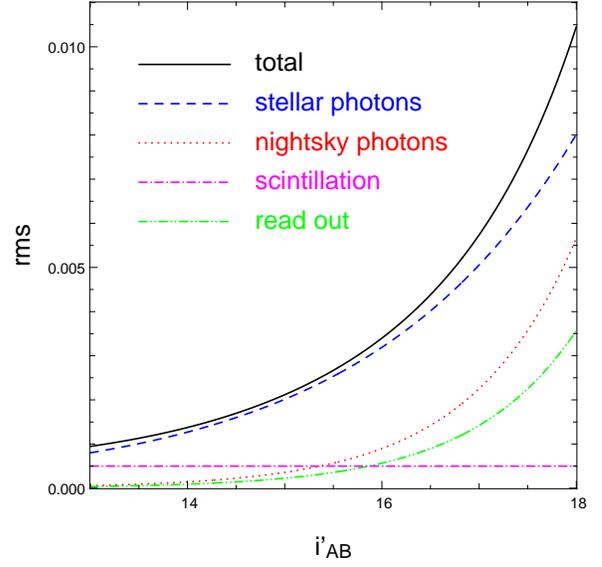}
	\caption{Total white noise and contributions of the four
    components.}
  \label{noise}
\end{figure}

For the white noise in our light curves we add four different Gaussian
components: stellar photon noise, sky background, readout and
scintillation noise. The photon noise of each star is estimated using
a preliminary exposure time calculator which has been calibrated by
observations taken during a pre-commissioning phase of the PanSTARRS1
telescope. We assume the sky background to be 20.15 mag per square
arcsecond which corresponds to a seven day distance to full moon. The
readout noise is assumed to be 8 $e^{-}$ per pixel. The scintillation
noise is estimated to be 0.5 mmag according to the formula of
\citet{1967AJ.....72..747Y,1993Obs...113...41Y} and is only of
importance at the very bright end of our target distribution.
\mbox{Fig. \ref{noise}} shows the white noise as function of
magnitude as well as the individual contributions.\\
For our calculations we assume a seeing of 1.2 arcsec, airmass of 1.4,
extinction coefficient of 0.08 and PSF fitting radius of 1.0 arcsec.
At the faint end (i' = 18 mag) the number of photons is on the order
of 15,500 for the object and 7,800 for the sky and therefore well
outside the Poisson statistics regime.
%
%
\subsection{Residual red noise model}
\label{subsec.red}
\begin{figure}
  \centering
  \includegraphics[width=0.4\textwidth]{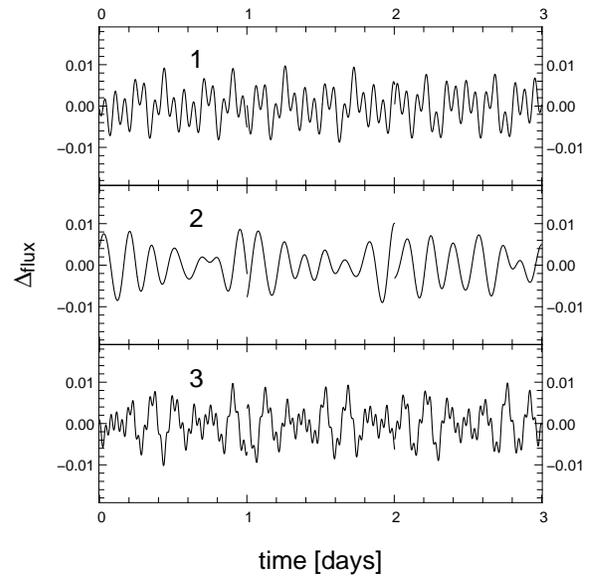}
	\caption{$\Delta_{flux}(t)$ for the fixed parameter models 1 to
    3. Because the random phase shift is calculated for each day
    individually there are discontinuities visible at integer day
    positions (local noon). In our simulation we calculate the phase
    shift for each observing block in order to ensure continuity
    within an observing block.}
  \label{red_lc}
\end{figure}

As detailed analyses of light curve datasets have shown, all transit
surveys suffer from non-Gaussian correlated noise sources, also known
as red noise. E.g. \cite{2006MNRAS.373..231P} analyzed the OGLE-III
light curves and calculated binned averages of subsets containing n
data points. They found the standard deviation $\sigma$ of these
averages can be parameterized to a good approximation by the
following formula :
\begin{equation}
  \sigma = \sqrt{\frac{\sigma_{white}^2}{n} + \sigma_{red}^2}
  \label{eq.red_noise}
\end{equation}
with $\sigma_{white}$ being the single point rms of the white noise
component and $\sigma_{red}^2$ being a constant red noise
contribution. With this equation one can model how the red noise
decreases the signal-to-noise ratio (S/N) of a transit light curve.\\
Application of algorithms to remove systematic effects, such as Sysrem
\citep{2005MNRAS.356.1466T} or TFA \citep{2005MNRAS.356..557K} have
been successfully applied by several groups resulting in a significant
reduction of the level of red noise
\citep[e.g.][]{2007A&A...476.1357S}. However, a small fraction of the
correlated noise always remained.\\
In our simulations we want to account for this residual red noise
(RRN). A simple model would be to increase the level of Gaussian noise
by a certain amount and therefore assume that the correlated nature is
of minor importance. For studies based only on S/N calculations one
could also use a parameterization like equation \ref{eq.red_noise}.
Since we are simulating light curves, we want to introduce a different
approach. We model the RRN by adding superimposed sine waves of
different wavelengths and amplitudes. This allows us to include the
effects the correlated noise has on the efficiency of the detection
algorithm, which could get confused by noise that is correlated on
timescales of a typical transit duration.\\\\
We add RRN according to the following model :
\begin{equation}
  \Delta_{flux}(t) = \sum\limits_{i} A_i \cdot sin( \frac{\pi}{\tau_i} t + p_{0,i} )
\end{equation}
with normalized amplitude $A_i$, timescale $\tau_i$ and random phase
shift $p_{0,i}$ of each component i. The phase shift is calculated for
each observing block independently and therefore changing with time
for a single light curve. This is done in order to avoid introducing
strong periodic signals that are coherent over a timescale longer than
a day.\\
For each model we start with relative amplitudes $A_i'$ which are
normalized in such a way that the rms of the added RRN
($rms_{red}$) is of value 1 mmag, 2 mmag, 3 mmag or 4 mmag :
\begin{equation}
  rms_{red} = \sqrt{\frac{\sum_{i} {A}_i^2}{2}}\quad .
\end{equation}
In order to analyze the influence of the timescales and amplitudes on
our results we construct a total of 9 different red noise models with
each of them having 3 or 4 components. Table \ref{red_noise} gives an
overview of the parameters of our red noise models. We refer to models
1 to 3 as 'fixed parameter' models because for these we select
arbitrary fixed values for $A_i'$ and $\tau_i$ which we use for all
light curve.  For models 4 to 9 we draw the relative amplitudes and
timescales randomly in a given range and for each light curve
individually. \mbox{Fig. \ref{red_lc}} shows $\Delta_{flux}(t)$ for
the fixed parameter models.
\begin{table*} 
  \scriptsize
  \setlength{\tabcolsep}{1mm}
  \centering
  \begin{tabular}{|c|c|c|c|c|c|c|c|c|c|c|c|c|} \hline
  model number  & $A_1'$       & $A_2'$       & $A_3'$       & $A_4'$       & $\tau_1$[min]    & $\tau_2$[min]    & $\tau_3$[min]    & $\tau_4$[min]    \\ \hline
  1             & 1            & 2            & 3            & 4            & 355              & 169              & 111              & 48               \\ \hline
  2             & 2            & 3            & 4            & 1            & 169              & 131              & 111              & 88               \\ \hline
  3             & 3            & 4            & 1            & 2            & 131              & 99               & 61               & 27               \\ \hline
  4             & random [1-4] & random [1-4] & random [1-4] & random [1-4] & random [300-400] & random [200-300] & random [100-200] & random [  0-100] \\ \hline
  5             & random [1-4] & random [1-4] & random [1-4] & random [1-4] & random [250-300] & random [200-250] & random [150-200] & random [100-150] \\ \hline
  6             & random [1-4] & random [1-4] & random [1-4] & random [1-4] & random [250-300] & random [200-250] & random [150-200] & ---              \\ \hline
  7             & random [1-4] & random [1-4] & random [1-4] & random [1-4] & random [250-300] & random [200-250] & ---              & random [100-150] \\ \hline
  8             & random [1-4] & random [1-4] & random [1-4] & random [1-4] & random [250-300] & ---              & random [150-200] & random [100-150] \\ \hline
  9             & random [1-4] & random [1-4] & random [1-4] & random [1-4] & ---              & random [200-250] & random [150-200] & random [100-150] \\ \hline
  \end{tabular}
  \caption{Dimensionless relative amplitudes $A_i'$ and 
    timescales $\tau_i$ of our different red noise models. Models 
    1 to 3 are fixed parameter models whereas for the others we 
    draw random values within a given range for each single 
    light curve individually. The timescales are chosen to cover 
    the range of expected transit durations.}
  \label{red_noise} 
\end{table*}
\subsection{Epochs of the observations}
\label{subsec.obsdates}
For each year the Pan-Planets survey has been granted a total of 120h
hours which will be executed in 1h or 3h blocks. Assuming a 33\% loss
due to bad weather we expect the data to be taken in 81 or 27 nights
per year depending on our survey strategy. The actual epochs of the
observations we use to construct our light curves are computed in the
following way. For each night in which our target field is visible for
at least 3h we calculate the range of visibility, namely the time the
field is higher than airmass 2 on the sky. This results in a 183 day
period starting on April 26th and ending on October 24th. We randomly
pick nights during the period of visibility and place the observing
block arbitrarily within the time span our target is higher than
airmass 2, as calculated earlier.\\
In our simulations we test 5 different scenarios with alternate
observations of 3 to 7 fields during the observing block. The time for
one exposure and readout is assumed to be 40s. Therefore the different
number of fields transform into cycle rates between 120s and
280s. Using the selected nights, the random position of a block within
a night and the cycle rate we construct a table of observational dates
which we use as input to the light curve simulations. For each
simulation run (which represents one possible outcome of the survey)
we draw new observational dates. Table \ref{obs_parameter} summarizes
the observational parameter depending on the survey strategy.
\begin{table*} 
  \scriptsize
  \setlength{\tabcolsep}{1mm}
  \centering
  \begin{tabular}{|c|c|c|c|c|} \hline
  \# of fields & block size & cycle rate & \# of data points & \# of data points \\
               &            &            & per night         & per year          \\ \hline\hline
  3            & 1h         & 120s       & 30                & 2430              \\ \hline
  4            & 1h         & 160s       & 23                & 1863              \\ \hline
  5            & 1h         & 200s       & 18                & 1458              \\ \hline
  6            & 1h         & 240s       & 15                & 1215              \\ \hline
  7            & 1h         & 280s       & 13                & 1053              \\ \hline
  3            & 3h         & 120s       & 90                & 2430              \\ \hline
  4            & 3h         & 160s       & 68                & 1863              \\ \hline
  5            & 3h         & 200s       & 54                & 1458              \\ \hline
  6            & 3h         & 240s       & 45                & 1215              \\ \hline
  7            & 3h         & 280s       & 39                & 1053              \\ \hline
  \end{tabular}								 
  \caption{Cycle rate and number of data points per night and year depending on observational strategy.}  
  \label{obs_parameter} 
\end{table*}
\subsection{Light curve analysis}
\label{subsec.analysis}
\begin{figure}
  \centering
  \includegraphics[width=0.4\textwidth]{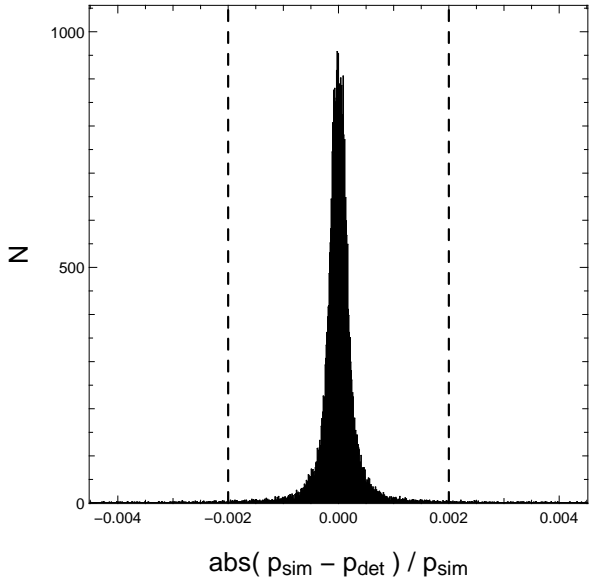}
	\caption{Deviation of the detected period $p_{det}$ from the
    simulated period $p_{sim}$ of a number of arbitrary selected
    observation runs. For a successful detection we require the
    detected period to deviate less than 0.2\% (dashed line).}
  \label{period_match}
\end{figure}

Each simulated light curve is analyzed by our detection algorithm
which is a box-fitting-least-squares (BLS) algorithm proposed by
\citet{2002A&A...391..369K}. The program folds the light curves with
trial periods in the range from 0.9 to 9.1 days and finds the best
$\chi^{2}$ fitting box corresponding to a fractional transit
length\footnote{the fractional transit length is defined as the
  transit duration divided by the period} $\tau$ between 0.01 and
0.1. For each detection the BLS algorithm provides period, S/N and the
number of individual transits. For a successful detection we require
the period found to match the simulated period within 0.2\% (see
\mbox{Fig. \ref{period_match}}). In addition, we impose the S/N to be
larger than 16 (see \S\ref{subsec.SNcut}) and the number of transits
to be at least equal to 3. The planet is also considered being
detected if the measured period is half or twice the simulated period
(to within 0.2\%). This can easily happen in case of unevenly sampled
light curves. For later analysis we store all input parameters of the
simulation and output parameters of the detection algorithm in a 
table.
\subsection{Signal-to-noise cut}
\label{subsec.SNcut}
To model a transit survey, it is very important to have a transparent
and reproducible procedure of applying cuts in the process of
selecting the candidates. The most important value is the minimum
S/N. The S/N of a transit light curve is defined as the transit depth
divided by the standard deviation of the photometric average of all
measurements taken during a transit. For a light curve with $N$
uniformly spaced data points with individual Gaussian error $\sigma$,
transit depth $\delta$ and a fractional transit length $\tau$ this 
is :
\begin{equation}
  S/N = \frac{\delta}{\sigma / \sqrt{N\tau}}\quad .
\end{equation}
In the presence of red noise the value of the S/N is reduced. Also the
actual shape of the transit, which is determined by limb-darkening and
ingress/egress, has an impact on the S/N. This effect is included
implicitly in our simulation.\\
Since the probability of finding a planet is small, the majority of
transit surveys use a low S/N cut of about 10. This results in a high
number of statistical and physical false
positives\footnote{statistical false positives are purely noise
  generated detections whereas physical false positives are true low
  amplitude variations (like e.g. in a blended binary system)} and has
made it necessary to include non-reproducible selection procedures
such as ''by-eye'' rejection. Pushing the S/N cut to the detection
limit makes it therefore difficult to model the detection
efficiency.\\
In the Pan-Planets survey we expect to find a very high number of
candidates already in the first year which will require a high amount
of radial velocity follow-up resources. The best candidates have the
highest S/N and will be followed-up first. We will most likely not be
able to follow-up all candidates down to the detection threshold of
$\sim$ 12 and therefore use a somewhat larger S/N cut. In this work we
calculate the expected number of detections using an S/N cut of 16.
\section{Results of the Monte Carlo simulations}
\label{results}
In this section we summarize the results of a total of 7.6 million
simulation runs. The computation time was 230\,000 CPU hours which we
distributed over a 486 CPU beowulf cluster.\\
In \S\ref{subsec.block_size} we show which block size (1h or 3h) is
more efficient for the Pan-Planets survey. In
\S\ref{subsec.rednoise_model} we compare the different RRN
models. Section \S\ref{subsec.number_fields} addresses the question of
the optimal number of fields (3 to 7). In \S\ref{subsec.final_result}
we summarize the actual number of VHJ, HJ, WJ, VHS we expect to find
using our preferred survey strategy. Finally, we show the results in
the case of observing the same fields during the second year of the
survey instead of monitoring new ones (\S\ref{subsec.2nd_year}). In
\S\ref{subsec.VHN} we study the potential to find Very Hot Neptunes
transiting M dwarfs.\\
Error estimates are only given for the final numbers in
\S\ref{subsec.final_result} and \S\ref{subsec.2nd_year}. All numbers
we present are scaled from the 1 sq.deg. Besan\c{c}on model to the
actual survey area of $N_{fields} \times$ 7 sq.deg assuming a constant
spectral type and magnitude distribution and a homogeneous density.\\
In order to check whether there are 7 fields of comparable density, we
count the total number of stars in the USNO-A2.0 catalog and compare
it to the total number of stars in the Besan\c{c}on model for a set of
different Galactic longitudes (Table \ref{number_count}). We assumed
an average color ($m_{USNO\_R}$-$m_{i'}$) of 0.25 mag. In the range
43.5 $\le$ l $\le$ 61.5 the number of stars in the Besan\c{c}on model
agrees well with the number of stars. The USNO density varies at a
level of 30\% with the average being $\sim$14\,000, close to the
density we assume in our simulations (l = 54.5). With a diameter of 3
deg. a total of 7 Pan-Starrs fields fit in this range.
\begin{table*} 
  \scriptsize
  \setlength{\tabcolsep}{1mm}
  \centering
  \begin{tabular}{|c|c||c|c|} \hline
    l      & b       & \# USNO-A2.0                     & \# Besan\c{c}on model   \\ 
    deg    & deg     & 13.25 $\le m_{USNO\_R} \le$ 16.25 & 13 $\le m_{i'} \le$ 16  \\ \hline\hline
    40.5   & -4.2    &  7748                            & 15103                   \\ \hline
    41.5   & -4.2    &  8439                            & 14623                   \\ \hline
    42.5   & -4.2    & 10670                            & 14352                   \\ \hline
    43.5   & -4.2    & 14814                            & 14248                   \\ \hline
    44.5   & -4.2    & 14248                            & 14208                   \\ \hline
    45.5   & -4.2    & 10906                            & 13754                   \\ \hline
    46.5   & -4.2    & 14910                            & 13645                   \\ \hline
    47.5   & -4.2    & 17018                            & 13194                   \\ \hline
    48.5   & -4.2    & 17065                            & 13175                   \\ \hline
    49.5   & -4.2    & 14482                            & 12959                   \\ \hline
    50.5   & -4.2    & 14295                            & 12370                   \\ \hline
    51.5   & -4.2    & 14424                            & 12459                   \\ \hline
    52.5   & -4.2    & 16737                            & 12260                   \\ \hline
    53.5   & -4.2    & 15890                            & 11997                   \\ \hline
    54.5   & -4.2    & 14131                            & 11770                   \\ \hline
    55.5   & -4.2    & 14555                            & 11705                   \\ \hline
    56.5   & -4.2    & 15682                            & 11456                   \\ \hline
    57.5   & -4.2    & 14562                            & 11370                   \\ \hline
    58.5   & -4.2    & 13195                            & 11058                   \\ \hline
    59.5   & -4.2    & 11301                            & 10877                   \\ \hline
    60.5   & -4.2    & 11194                            & 10436                   \\ \hline
    61.5   & -4.2    &  9188                            & 10489                   \\ \hline
    62.5   & -4.2    &  6181                            & 10139                   \\ \hline
    63.5   & -4.2    &  4968                            &  9903                   \\ \hline
  \end{tabular}
  \caption{Total number of stars per sq. deg. according to the 
    USNO-A2.0 catalog and the Besan\c{c}on model.}
  \label{number_count}
\end{table*}
\subsection{Influence of the size of the observing blocks}
\label{subsec.block_size}
We investigate the influence of the observing block size on the number
of detections in the Pan-Planets survey. Table \ref{results_block_1yr}
lists the average number of VHJ and HJ found with 1h and 3h blocks
after the application of our detection cuts, as described in
\S\ref{subsec.SNcut}.\\
The first three columns list the planet population and the survey
strategy (i.e. number of fields and observing block size). The fourth
column shows the average numbers of all simulated transiting planet
light curves having an S/N of 16 or more (without requiring 3 transits
and without running the detection algorithm). Here the numbers are
very similar comparing the 1h to the 3h block strategies.\\
To understand this, one has to consider that a planet spends a certain
fraction of its orbit in transit phase (also known as fractional
transit length $\tau$). This fraction depends mainly on the
inclination and period of the orbit as well as the radius of the host
star. For a given $\tau$ the average number of points in transit ($N$
$\cdot$ $\tau$) only depends on the total number of observations $N$
and is therefore independent of the block size. The same applies to
the S/N which, for fixed transit depth and photometric noise
properties, depends only on the number of points in transit to a good
approximation. Therefore, if only a minimum S/N is required, the
number of detections is comparable for a strategy with 1h blocks and
with 3h blocks, with minor differences arising from limb-darkening and
ingress/egress effects.\\
Although the number of points in transit is the same for both
strategies, one 3h block covers on average a bigger part of the
transit compared to a 1h block. As a consequence the average number of
individual transits must be lower in the case of 3h blocks. If we
impose the additional cut of requiring at least 3 transits to be
visible in the light curve (column 5), the expected number of planets
found is lower for the 3h blocks compared to the 1h blocks. With a 3h
block strategy the number of light curves passing the S/N cut and
having 3 or more transits is on average 53\% lower for HJ and 26\%
lower for VHJ. For the longer period HJ this effect is stronger due to
the fact that the number of visible transits is lower in general.\\
In order the planet to be considered detected (as described in
\S\ref{simulations}), we not only require S/N $\ge$ 16 and at least 3
visible transits, but also that the BLS detection algorithm finds the
correct period (allowing for twice and half the correct value). The
impact of this additional selection cut is shown in columns 6-10 for
light curves with 0 mmag, 1 mmag, 2 mmag, 3 mmag and 4 mmag
RRN\footnote{we restrict ourself here to RRN of model 4 - our favored
  model (see \S\ref{subsec.rednoise_model})}. Without RRN, most
planets are found by the BLS algorithm. The loss is marginally higher
in the case of 3h blocks which is a consequence of the generally lower
number of transits, since the BLS algorithm is more efficient if more
transits are present. Comparing the results for 1h and 3h blocks we
find that in case the RRN level is 2 mmag, the number of detected
planets without RRN is on average 59\% lower for HJ and 30\% lower for
VHJ in the 3h block case.\\
Including RRN, fewer planets are detected by the BLS algorithm and the
discrepancy between 1h and 3h blocks increases. For a typical RRN
level of 2 mmag we find on average 71\% less HJ and 45\% less VHJ with
3h blocks compared to 1h blocks.\\
As an additional test we perform the same analysis for a campaign with
twice the amount of observing time spread over 2 years. This would
correspond to a strategy where we stay on the same target fields in
the second year of the Pan-Planets survey. Also in this case, 1h
blocks are more efficient than 3h blocks. Assuming the RRN level is 2
mmag, we find that the number of detected planets is on average 34\%
lower for HJ and 59\% lower for VHJ in the 3h block case. The details
of the simulations for a 2 yr campaign can be found in
\S\ref{subsec.2nd_year}. In the following we restrict our results to
1h blocks.
\begin{table*}  
  \scriptsize
  \setlength{\tabcolsep}{1mm}
  \centering
  \begin{tabular}{|c|c|c||c|c|c|c|c|c|c|} \hline
    population & \# of fields & block & S/N $\ge$ 16 & $\ge$ 3 transits & 0 mmag & 1 mmag & 2 mmag & 3 mmag & 4 mmag  \\ \hline\hline
    VHJ        & 3            & 1h    & 13.13        & 11.39            & 10.95  & 10.18  &  7.73  &  5.33  &  3.61   \\ \hline
    VHJ        & 4            & 1h    & 16.09        & 13.91            & 13.37  & 12.54  &  9.82  &  7.00  &  4.82   \\ \hline
    VHJ        & 5            & 1h    & 18.53        & 16.12            & 15.52  & 15.06  & 12.06  &  8.66  &  6.10   \\ \hline
    VHJ        & 6            & 1h    & 20.67        & 18.00            & 17.34  & 16.75  & 14.00  & 10.17  &  7.38   \\ \hline
    VHJ        & 7            & 1h    & 22.44        & 19.57            & 18.87  & 18.44  & 15.76  & 11.72  &  8.44   \\ \hline\hline
    VHJ        & 3            & 3h    & 12.50        &  8.16            &  7.50  &  6.45  &  4.04  &  2.57  &  1.61   \\ \hline
    VHJ        & 4            & 3h    & 15.04        &  9.95            &  9.16  &  8.32  &  5.32  &  3.39  &  2.15   \\ \hline
    VHJ        & 5            & 3h    & 17.69        & 11.88            & 10.95  &  9.66  &  6.48  &  4.23  &  2.75   \\ \hline
    VHJ        & 6            & 3h    & 20.09        & 13.61            & 12.54  & 11.01  &  7.84  &  5.23  &  3.21   \\ \hline
    VHJ        & 7            & 3h    & 21.43        & 14.64            & 13.47  & 12.51  &  9.03  &  6.09  &  3.84   \\ \hline\hline
     HJ        & 3            & 1h    & 15.93        & 10.83            &  9.75  &  8.65  &  5.52  &  3.36  &  2.06   \\ \hline
     HJ        & 4            & 1h    & 18.31        & 12.11            & 10.87  & 10.07  &  6.97  &  4.37  &  2.74   \\ \hline
     HJ        & 5            & 1h    & 20.54        & 13.69            & 12.22  & 11.40  &  8.43  &  5.30  &  3.37   \\ \hline
     HJ        & 6            & 1h    & 22.45        & 14.82            & 13.22  & 12.32  &  9.41  &  6.15  &  4.07   \\ \hline
     HJ        & 7            & 1h    & 23.60        & 15.54            & 13.78  & 13.16  & 10.28  &  6.86  &  4.45   \\ \hline\hline
     HJ        & 3            & 3h    & 14.06        &  4.98            &  3.91  &  2.91  &  1.49  &  0.85  &  0.53   \\ \hline
     HJ        & 4            & 3h    & 16.30        &  5.78            &  4.49  &  3.54  &  1.88  &  1.04  &  0.66   \\ \hline
     HJ        & 5            & 3h    & 18.22        &  6.45            &  4.97  &  4.09  &  2.31  &  1.34  &  0.84   \\ \hline
     HJ        & 6            & 3h    & 20.06        &  7.04            &  5.38  &  4.49  &  2.74  &  1.56  &  0.98   \\ \hline
     HJ        & 7            & 3h    & 21.59        &  7.60            &  5.73  &  5.07  &  3.20  &  1.89  &  1.15   \\ \hline
  \end{tabular}                 
  \caption{Influence of the block size shown on the basis of the 
    number of planets detected in a 1 yr campaign after applying 
    different detection cuts.}
  \label{results_block_1yr}
\end{table*}
\subsection{Influence of the residual red noise model}
\label{subsec.rednoise_model}
In this section we compare the results of nine different RRN models
which have been introduced in \S\ref{subsec.red}. In addition, we
compare the red noise models to a scenario where we add additional
uncorrelated white noise by the same amount as the RRN level. Table
\ref{results_red} shows the number of HJ and VHJ found with a 1h block
strategy and 2 mmag RRN for each of the 9 different red noise models,
as well as for the increased white noise model.\\
In general the increased white noise model results in a significantly
higher number of detections compared to the RRN models (on average
22\% and 39\% higher for VHJ and HJ respectively). This shows that the
effect of the RRN on the efficiency of the BLS algorithm is strong and
needs to be taken into account in our simulations.\\
Comparing the individual RRN models to each other we find that for the
fixed parameter models (1 to 3) the number of detections is 8\% and
13\% higher for VHJ and HJ respectively than for the random models (4
to 9). The individual results of the random models are all very
similar and vary only by a few percent. In the following we restrict
our results to the red noise model 4, since it is the most general of
all models with 4 components and random timescales ranging from 0 to 
400 minutes.
\begin{table*} 
  \scriptsize
  \setlength{\tabcolsep}{1mm}
  \centering
  \begin{tabular}{|c|c||c|c|c|c|c|c|c|c|c||c|} \hline
    population & \# of fields & model 1 & model 2 & model 3 & model 4 & model 5 & model 6 & model 7 & model 8 & model 9 & white   \\ \hline\hline
    VHJ        & 3            &  9.10   &  8.26   &  8.72   &  7.73   &  7.90   &  7.80   &  7.84   &  7.85   &  7.75   & 10.91   \\ \hline
    VHJ        & 4            & 11.30   & 10.31   & 11.17   &  9.82   &  9.81   &  9.88   &  9.92   &  9.97   &  9.83   & 13.30   \\ \hline
    VHJ        & 5            & 13.53   & 12.66   & 13.34   & 12.06   & 11.95   & 12.29   & 12.43   & 12.23   & 12.08   & 15.08   \\ \hline
    VHJ        & 6            & 15.81   & 14.66   & 15.01   & 14.00   & 14.16   & 14.12   & 14.01   & 13.89   & 14.23   & 17.11   \\ \hline
    VHJ        & 7            & 17.24   & 16.13   & 16.97   & 15.76   & 15.77   & 15.95   & 15.85   & 15.74   & 15.81   & 18.44   \\ \hline
     HJ        & 3            &  7.04   &  6.01   &  6.74   &  5.52   &  5.63   &  5.69   &  5.61   &  5.54   &  5.60   &  9.39   \\ \hline
     HJ        & 4            &  8.51   &  7.37   &  7.99   &  6.97   &  6.80   &  6.91   &  6.89   &  6.91   &  7.00   & 10.51   \\ \hline
     HJ        & 5            &  9.84   &  8.83   &  9.75   &  8.43   &  8.32   &  8.29   &  8.33   &  8.32   &  8.45   & 12.10   \\ \hline
     HJ        & 6            & 11.13   & 10.08   & 10.53   &  9.41   &  9.42   &  9.41   &  9.37   &  9.51   &  9.41   & 13.15   \\ \hline
     HJ        & 7            & 11.77   & 10.78   & 11.37   & 10.28   & 10.39   & 10.39   & 10.42   & 10.31   & 10.31   & 13.64   \\ \hline
  \end{tabular}
  \caption{Influence of the red noise model for 1h blocks and a 
    RRN level of 2 mmag.}
  \label{results_red} 
\end{table*}
\subsection{Influence of the number of fields}
\label{subsec.number_fields}
\begin{figure}
  \centering
  \includegraphics[width=0.4\textwidth]{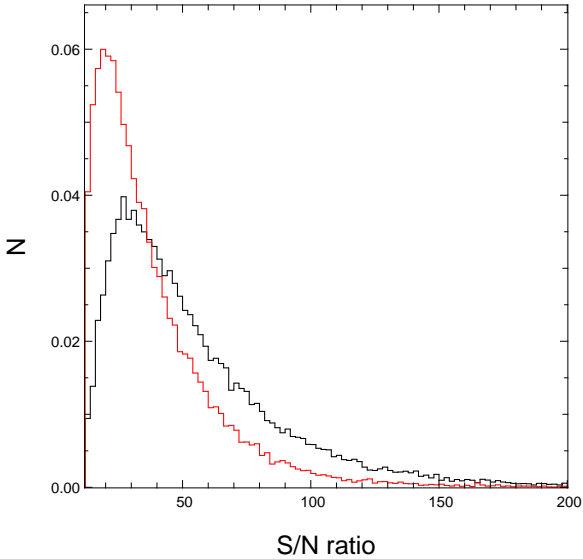}
  \caption{Normalized S/N distribution of VHJ detections for a 3 field
    strategy (black) and a 7 field strategy (red) assuming 1h blocks
    and 2 mmag RRN (model 4).}
  \label{results_SN_hist}
\end{figure}

In order to optimize the survey with respect to the number of
alternating fields monitored during an observing block, we compare the
number of detections for each of the 5 strategies (3 to 7 fields). We
do not test more than 7 fields, because it is not sure if we can find
a higher number of fields with comparable density (see
\S\ref{results}).  Note also, that with more than 7 fields, the number
of data points per light curve would be less than 1\,000 and the cycle
rate longer than 5 minutes which would complicate the process of
eliminating false positives on the basis of the light curve
shape\footnote{it is important to well sample the ingress/egress part
  of the transits which has a duration of approximately 15-20
  minutes}. We limit our simulations to the above selected 1h blocks
(see \S\ref{subsec.block_size}) and 2 mmag RRN of model 4 (see
\S\ref{subsec.rednoise_model}).\\
In general, the total number of detections depends on the number of
fields in two counteracting ways: on the one hand, observing more
fields results in more target stars and therefore more transiting
planet systems that can be detected; on the other hand, observing more
fields results in a lower number of data points per light curve and
thus the S/N of each transit candidate is shifted to a lower value.
The latter effect is stronger for faint stars because the S/N is
generally lower whereas for brighter stars the S/N is high enough in
most cases.\\
The number of detections in the first year of the Pan-Planets survey
for different number of fields is shown in Table
\ref{results_fields}. For all planet populations (i.e. VHJ, HJ, WJ and
VHS) we find more planets with a higher number of fields. The loss in
S/N is over-compensated by the higher number of target stars. In
Fig. \ref{results_SN_hist} we show the S/N distributions of VHJ
detections for a 3 field and a 7 field strategy. The S/N distribution
of VHJ peaks at a higher level than our cut of 16, even for the 7
field strategy, which explains why observing more fields results in
more detections.\\
In the case of a 2 yr campaign the situation is the same. For all four
planet populations it is more efficient to observe a higher number of
fields (see Table \ref{results_fields_2yr}). Therefore we conclude
that observing 7 fields is the most efficient strategy and
restrict our results in the following to 7 fields.
\begin{table*}
  \scriptsize
  \setlength{\tabcolsep}{1mm}
  \centering
  \begin{tabular}{|c||c|c|c|c|} \hline
    \# fields & VHJ   & HJ    & WJ   & VHS   \\ \hline\hline
    3         &  7.73 &  5.52 & 1.60 & 2.26  \\ \hline
    4         &  9.82 &  6.97 & 1.95 & 2.63  \\ \hline
    5         & 12.06 &  8.43 & 2.44 & 3.05  \\ \hline
    6         & 14.00 &  9.41 & 2.61 & 3.40  \\ \hline
    7         & 15.76 & 10.28 & 2.78 & 3.51  \\ \hline
  \end{tabular}
  \caption{Number of planets found in the first year depending on the number 
    of target fields, assuming 1h blocks and 2 mmag RRN (model 4).}
  \label{results_fields}
\end{table*}
\begin{table*}
  \scriptsize
  \setlength{\tabcolsep}{1mm}
  \centering
  \begin{tabular}{|c||c|c|c|c|} \hline
    \# fields & VHJ   & HJ    & WJ   & VHS   \\ \hline\hline
    3         & 11.16 & 11.90 & 4.85 & 3.95  \\ \hline
    4         & 14.72 & 15.21 & 6.18 & 5.07  \\ \hline
    5         & 17.96 & 18.34 & 7.46 & 5.94  \\ \hline
    6         & 21.25 & 21.22 & 8.66 & 6.68  \\ \hline
    7         & 24.13 & 23.55 & 9.48 & 7.49  \\ \hline
  \end{tabular}
  \caption{Number of planets found in 2 yr depending on the number of target 
    fields, assuming 1h blocks and 2 mmag RRN (model 4).}
  \label{results_fields_2yr}
\end{table*}
\subsection{The expected number of planets in the Pan-Planets survey}
\label{subsec.final_result}
\begin{figure}
  \centering
  \includegraphics[width=0.4\textwidth]{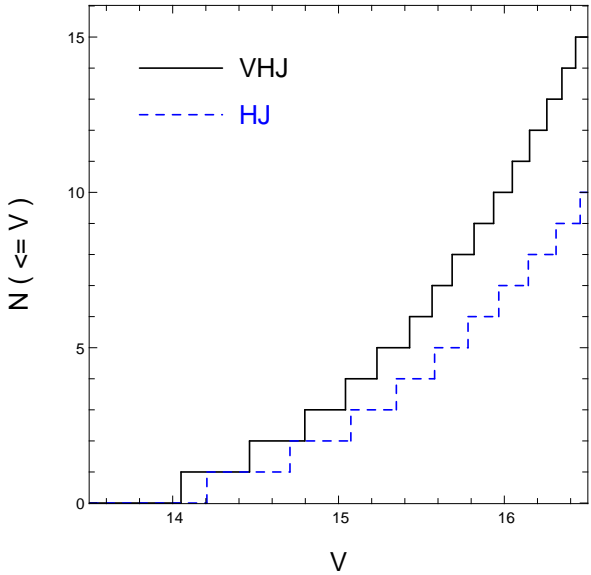}
  \caption{Cumulative host star brightness distributions for each
    planet population for a 1 yr campaign with 2 mmag RRN (model 4).}
  \label{results_mag}
\end{figure}

In the previous sections we have identified our preferred survey
strategy with 1h blocks and alternating among 7 fields. In addition we
selected RRN model 4 as our preferred one. For these parameters we
performed more detailed simulations in order to calculate the expected
number of detections of the Pan-Planets project (including error
estimates) and to study the parameter distributions of the detected
planets in detail. For each of 4 different RRN levels (1 mmag, 2 mmag,
3 mmag and 4 mmag) and 4 planet populations (VHJ, HJ, WJ and VHS) we
performed 25\,000 simulation runs. The number of detections depending
on the level of RRN are shown in \mbox{Table \ref{final_results_1yr}}.
Fig. \ref{results_mag} shows the cumulative distribution of the host
star brightness for each planet populations for 2 mmag of RRN (model
4).\\
Our predicted numbers are affected by two sources of uncertainties.
The first and dominant one is the uncertainty of the planet frequency
taken from \citet{2006AcA....56....1G} which is caused by the low
number statistics of the OGLE detections. This uncertainty is not
included in Table \ref{final_results_1yr} and must be taken into
account by scaling all of our HJ results by a factor of
$1^{+1.37}_{-0.59}$ and all of our VHJ results by a factor of
$1^{+1.10}_{-0.54}$, as published in \citet{2006AcA....56....1G}.\\
The second uncertainty is a direct result of our simulations. Each
simulation run represents one possible outcome of \mbox{1 sq.deg.} of
the Pan-Planets survey. Since the simulated observational epochs
change from one run to the other and since in each run different stars
are attributed to planets with different orbital parameters, an
intrinsic scatter in the number of planets in found in each run. The
combination of 49 randomly chosen runs represents one possible outcome
of the full 49 sq.deg. survey (7 fields). From the histogram of these
combinations we derive 68\% confidence intervals for our predicted 
numbers.
\begin{figure}
  \centering
  \includegraphics[width=0.4\textwidth]{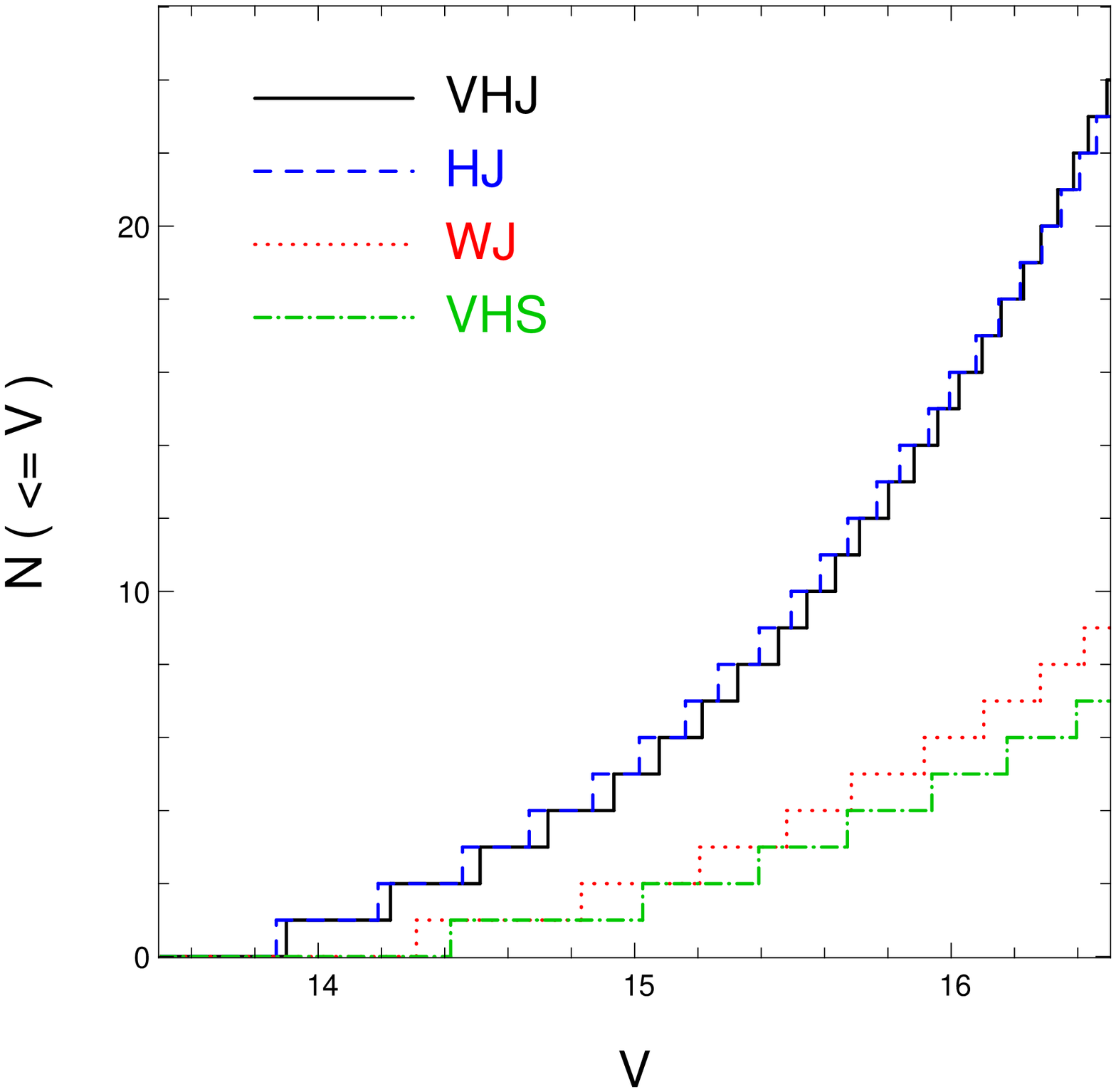}
  \caption{Cumulative host star brightness distributions for each
    planet population for a 2 yr campaign with 2 mmag RRN (model 4).}
  \label{results_mag_2yr}
\end{figure}
\begin{table*}
  \scriptsize
  \setlength{\tabcolsep}{1mm}
  \centering
  \begin{tabular}{|c||c|c|c|c|} \hline
    RRN level & VHJ            & HJ             & WJ                   & VHS                  \\ \hline\hline
    1 mmag    & $18.4 \pm 4.3$ & $13.2 \pm 3.7$ & $3.8 \pm 2.0$        & $5.2 \pm 2.3$        \\ \hline
    2 mmag    & $15.8 \pm 4.0$ & $10.3 \pm 3.2$ & $2.8 \pm 1.7$        & $3.5 \pm 1.9$        \\ \hline
    3 mmag    & $11.7 \pm 3.5$ & $ 6.9 \pm 2.7$ & $1.8 ^{+1.4 }_{-1.3 }$ & $1.9 \pm 1.4$        \\ \hline
    4 mmag    & $ 8.4 \pm 2.9$ & $ 4.5 \pm 2.1$ & $1.1 ^{+1.1 }_{-0.9 }$ & $0.9 ^{+1.0 }_{-0.7 }$ \\ \hline
  \end{tabular}
  \caption{Number of planets found in the first year for our selected 
    survey strategy as a function of residual red noise level 
    (model 4).}
  \label{final_results_1yr} 
\end{table*}
\subsection{Number of expected planets in a two year campaign}
\label{subsec.2nd_year}
\begin{figure}
  \centering
  \includegraphics[width=0.4\textwidth]{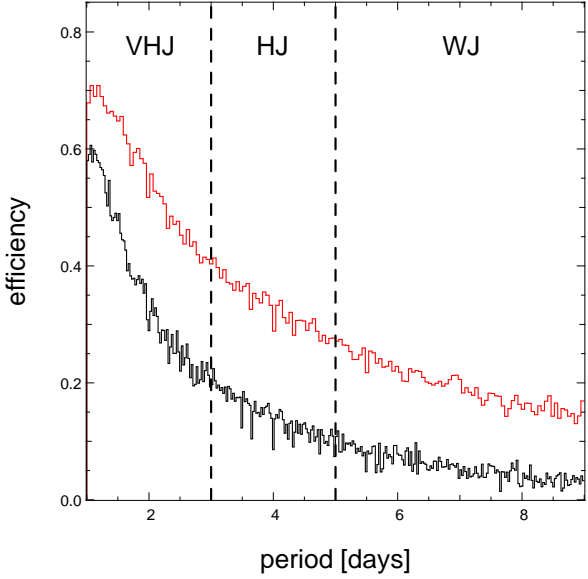}
  \caption{Fraction of all transiting VHJ, HJ and WJ that are detected
    as a function of period for a 1 yr campaign (lower black line) and
    a 2 yr campaign (upper red line).}
  \label{OPC}
\end{figure}

The Pan-Planets project has a lifetime of 3.5 years. In the previous
sections we focus mainly on the first year of the survey. In this
section we show the results of our simulations for a 2 year
campaign. In particular, we address the question whether the project
is more efficient if we stay on the same fields or if we choose new
targets assuming that we find fields with similar densities (see
\S\ref{conclusions}).\\
Table \ref{final_results_2yr} shows the number of planets detected in
a 2 yr campaign for four different levels of RRN. Except for VHJ, we
more than double the number of detections for each planet
population. For the longer period WJ the gain is a factor of 3. In a 1
yr campaign most of these planets show less than 3 transits and are
not detected. Adding the observations of the second year, the number
of transits increases and many of the previously undetected planets
are found.\\
In addition, staying on the same fields in the second year increases
the S/N of all transit light curves, due to the higher number of data
points taken during a transit. Planets that have an insufficiently
high S/N after the first year are detected after the second year.
This is particularly true for VHS.\\
Fig. \ref{OPC} shows the fraction of all transiting Jupiter-sized
planets (VHJ, HJ, WJ) that are detected in the first year of the
survey (lower black line) and in the 2 yr campaign (upper red line).
In the first year, the average efficiencies are 26.3\%, 10.6\% and
4.3\% for VHJ, HJ and WJ respectively.  Planets that have been missed
do not have the required S/N, show less than 3 transits in the light
curve, or the BLS algorithm found a wrong period. Extending the survey
to the second year increases the efficiency significantly to 39.8\%, 
24.0\% and 14.3\% for VHJ, HJ and WJ respectively.
\begin{table*}
  \scriptsize
  \setlength{\tabcolsep}{1mm}
  \centering
  \begin{tabular}{|c||c|c|c|c|} \hline
    RRN level & VHJ            & HJ             & WJ             & VHS             \\ \hline\hline
    1 mmag    & 28.1 $\pm$ 5.3 & 29.3 $\pm$ 5.4 & 12.4 $\pm$ 3.5 & 10.6 $\pm$ 3.3  \\ \hline
    2 mmag    & 24.1 $\pm$ 4.9 & 23.6 $\pm$ 4.9 &  9.5 $\pm$ 3.1 &  7.5 $\pm$ 2.7  \\ \hline
    3 mmag    & 19.0 $\pm$ 4.4 & 16.5 $\pm$ 4.1 &  6.1 $\pm$ 2.5 &  4.0 $\pm$ 2.1  \\ \hline
    4 mmag    & 14.4 $\pm$ 3.8 & 10.9 $\pm$ 3.3 &  4.0 $\pm$ 2.0 &  2.1 $\pm$ 1.5  \\ \hline
  \end{tabular}
  \caption{Number of planets found with 160h hours of observations in 
    2 yrs for our selected survey strategy as a function of residual 
    red noise level (model 4).}
  \label{final_results_2yr} 
\end{table*}
\subsection{The detection of Very Hot Neptunes}
\label{subsec.VHN}
\begin{figure}
  \centering
  \includegraphics[width=0.4\textwidth]{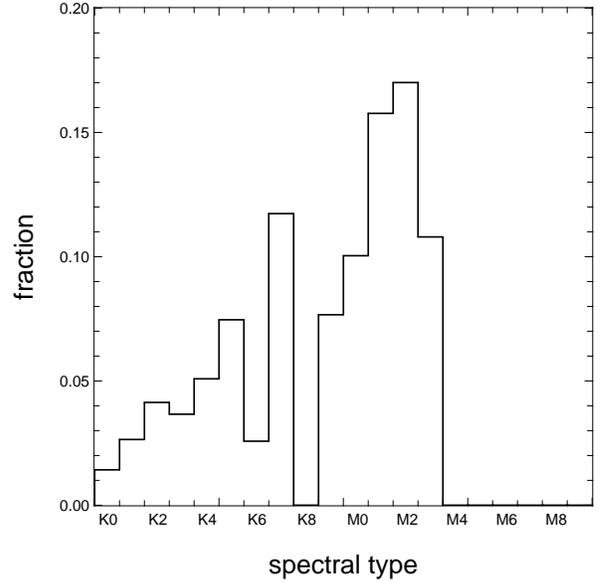}
  \caption{Host star spectral type distribution of all detected VHN
    for a 2 yr campaign with 2 mmag RRN (model 4). Pan-Planets is
    sensitive to VHN transiting late K and early M dwarfs.}
  \label{M-dwarfs}
\end{figure}

\begin{figure}
  \centering
  \includegraphics[width=0.4\textwidth]{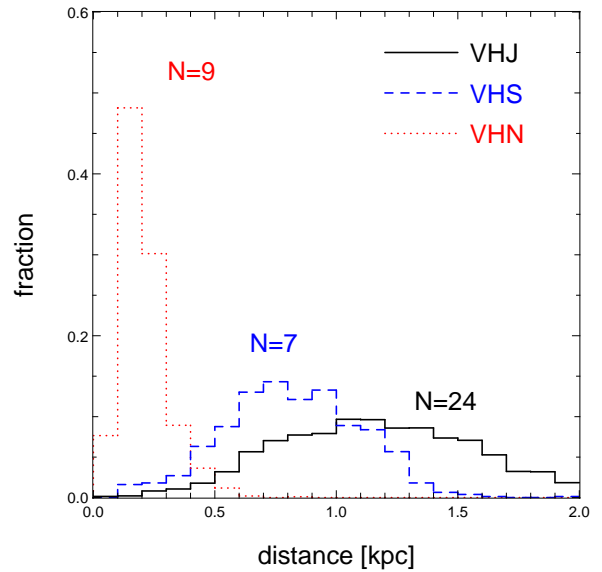}
  \caption{Distance distribution of the detected VHJ, VHS and VHN for
    a 2 yr campaign with 2 mmag RRN (model 4). Due to the lower
    transit depth smaller planets can only be detected around closer
    stars.}
  \label{distance}
\end{figure}

In this section we study the potential to find Very Hot Neptunes
transiting M dwarfs. The radius ratio between planet and star is much
higher for low mass stars which results in much deeper transits and
therefore a higher detection probability. According to the
Besan\c{c}on model there are a total of 34\,000 M dwarfs brighter than
AB-magnitude $m_{i'}$ = 18 mag in 7 fields of 7 sq.deg. each. These
objects are particularly interesting, since the composition of planets
in this mass range is rather unknown (gaseous, icy or rocky). Also the
habitable zone is much closer to the star due to its lower surface
temperature. Note, that only one planet transiting an M dwarf has been
detected so
far.\\
We consider all transiting VHN candidates down to host star
brightnesses of $m_{i'}$ = 18 mag to be interesting objects, although
the spectroscopical follow-up will be very challenging. New high
resolution near infrared spectrographs will help to confirm
these very red objects.\\
To study the potential of Pan-Planets to find transiting VHN we
perform simulations for the whole input stellar distribution and
analyze the spectral type distribution of the hosts stars of all
successful detections (Fig. \ref{M-dwarfs}). The Pan-Planets survey is
sensitive to close-in Neptune-sized planets around late K and early M
dwarfs if the frequency of these stars hosting Neptunes is as large as
5\%. The number of VHN detections after the first and the second year
is listed in Table \ref{VHN_results} for 4 different residual rednoise
levels. Assuming 2 mmag of RRN we expect to find 3
VHN after the first and 9 VHN after the second year.\\
Further we analyze the distance distributions of all detected VHJ, VHS
and VHN systems (Fig. \ref{distance}). The volume probed strongly
depends on the radius of the planet. For lower mass radius the transit
depth is generally smaller and therefore the photometric precision
needed to detect the transits must be higher, which is only the case
for closer and thus brighter systems. Note that for HJ and WJ the
distance distributions are very similar to the VHJ
distribution.
\begin{table*}
  \scriptsize
  \setlength{\tabcolsep}{1mm}
  \centering
  \begin{tabular}{|c||c|c|} \hline
    RRN level & 1 yr          & 2 yr            \\ \hline\hline
    1 mmag    & $5.7 \pm 2.4$ & $16.6 \pm 4.1$  \\ \hline
    2 mmag    & $3.5 \pm 1.9$ &  $9.7 \pm 3.1$  \\ \hline
    3 mmag    & $2.4 \pm 1.6$ &  $6.3 \pm 2.5$  \\ \hline
    4 mmag    & $1.7 \pm 1.4$ &  $4.6 \pm 2.2$  \\ \hline
  \end{tabular}
  \caption{Number of VHN detections after the first and second year of the 
    Pan-Planets survey for our selected survey strategy for different residual 
    red noise levels (model 4).}
  \label{VHN_results} 
\end{table*}
\section{Consistency check with the OGLE-III survey}
\label{consistency}
\citet{2006AcA....56....1G} have modeled the OGLE-III survey in order
to derive absolute frequencies of HJ and VHJ. In our simulations we
are using these frequencies to predict the number of detections for
the Pan-Planets survey. In order to verify our results we performed a
consistency check by modeling the OGLE-III survey and comparing the
results to the actual number of planets found. We limit this test to
the 3 Carina fields where 3 planets have been found (1 HJ, 2 VHJ). The
2 bulge fields that have also been observed during the OGLE-III
campaign are more difficult to model due to a stronger blending and a
higher uncertainty in the input stellar distribution.\\
We obtained the Besan\c{c}on model population of the 3 Carina fields
(CAR100, CAR104, CAR105) for stars in the magnitude range of 13.7 $\le
I_{mag} \le$ 17.0. The overall noise level as a function of magnitude
has been determined by \citet{2006AcA....56....1G} to be :
\begin{equation}
   \sigma = -0.723 + 0.1544 \cdot I_{mag} - 0.01094 \cdot I_{mag}^2 + 0.000259 \cdot I_{mag}^3 . 
\end{equation}
In their simulations \citet{2006AcA....56....1G} did not include
correlated noise sources, instead they account for systematics by
using an increased S/N cut. In order to be as consistent as possible
we follow the same procedure and do not split the over-all noise in
red and white noise components (as we did in the Pan-Planets
simulations). We use the radius and period distributions for HJ and
VHJ introduced in \S\ref{subsec.planets}. The epochs of the
observations were taken from the light curve of
OGLE-TR-74\footnote{one of the OGLE-III candidates} with 1,200 epochs
taken from February to May 2002.\\
After simulating the light curves in the same way as described in
\S\ref{simulations} we run the BLS algorithm and check for a correct
period recovery. In addition we apply the following cuts which have
been used by the OGLE group and are summarized in detail in Section 3
and Table 1 in \citet{2006AcA....56....1G}: the transit depth $\delta$
must be smaller than 0.04 mag ($\sim$3.62\%); the S/N greater than
11.6; the signal detection efficiency\footnote{quality parameter
  provided by the BLS algorithm for each detection} larger than 3.8;
the number of transits is required to be at least 3; and finally, the
color $(V-I)_0$ must be greater than 0.4. Note that we have not
imposed any cut on the transit depth in our simulations for the
Pan-Planets survey since a Jupiter-sized planet transiting an M dwarf
can have a fairly high transit depth. Further we do not use a color
since in our simulations we include only late type dwarfs a
priori.\\
In total we simulated 50\,000 runs for each of the five planet
populations. On average we find 2.18 VHJ and 1.46 HJ which is in
reasonable good agreement with the actual number of 2 VHJ and 1 HJ
found by the OGLE group.\\
According to our simulations the OGLE-III carina survey was not
sensitive to one of the other 3 planet populations we tested. We find
on average 0.45 WJ, 0.12 VHS and zero VHN which is agreement with 
none being found by OGLE.
\section{Conclusion}
\label{conclusions}
The aim of this work was to study the influence of the survey strategy
on the efficiency of the Pan-Planets project and to predict the number
of detections for an optimized strategy.\\
Our calculations are based on the simulation of realistic light curves
including the effects of limb-darkening, ingress/egress and
observational window functions. In addition we have introduced a model
to simulate correlated (red) noise which allows us to include the
effects of correlated noise on the efficiency of the BLS detection
algorithm. Our approach can be applied to any transit survey as
well.\\\\
Below we summarize the caveats and assumptions that were made in our
simulations :
\begin{itemize}
\item{Our results depend on the spectral type and magnitude
    distribution of the Besan\c{c}on model. The model does not include
    second order substructure such as spiral arms.}
\item{We neglect the effects of blending. Due to crowding into the
    direction of the Galactic disk some stars are blended by
    neighboring sources.}
\item{We assume all planets that are detected by the BLS-algorithm to
    be followed-up and confirmed spectroscopically. In particular, we
    assume that no true candidate is rejected by any candidate
    selection process. The detailed follow-up strategy of the
    Pan-Planets survey will be presented in Afonso et al. (in prep.).}
\item{Our simulations are done for 1 sq.deg. and the results are
    scaled to the actual survey area. We assume that all fields (3, 4,
    5, 6 or 7 case) have homogeneous densities and non-varying (or
    similar) stellar populations. Simple number counts on the
    USNO-catalog showed that we can find up to 7 fields with similar
    total number of stars (see \S\ref{results}). For a larger number
    of fields, the assumption of a constant density might be too
    optimistic since we are restricted to fields that are close to
    each other in order to keep the observational overhead low.}
\item{Our results directly scale with the assumed planet frequencies.
    The values of 0.14\% and 0.31\% we use for VHJ and HJ have
    uncertainties of a factor of 2. For WJ, VHS and VHN we have used
    hypothetical values of 0.31\%, 0.14\% and 5\% respectively.  After
    completion of the Pan-Planets survey we will be able to derive
    more accurate absolute frequencies for all five planet
    populations.}
\item{The quality of the data is assumed to be homogeneously good over
    the whole detector area. Bad pixel regions and gaps between the
    individual CCDs are not taken into account and result in an
    effective field of view that is smaller than 7 sq.deg.}
\end{itemize}
Comparing different observing strategies we found that observing more
fields is more efficient. Concerning the observation time per night,
we compared 1h blocks to 3h blocks and found the shorter ones
to be more efficient. This is still the case for a 2 yr campaign.\\\\
For an RRN level of 2 mmag we expect to find up to 15 VHJ and 10 HJ in
the first year around stars brighter than V = 16.5 mag. The survey
will also be sensitive to planets with longer periods (WJ) and smaller
radii (VHS and VHN). Assuming that the frequencies of stars with WJ
and VHS is 0.31\% and 0.14\% respectively, we expect to find up to
2 WJ and 3 VHS in the same magnitude range.\\
We found that observing the same fields in the second year of the 3.5
yr lifetime of the survey is more efficient than choosing new fields.
We expect to find up to 24 VHJ, 23 HJ, 9 WJ and 7 VHS. In particular
for longer periods (HJ and WJ) and smaller radii (VHS) we will more
than double the number of detections of the first year if we continue
to observe the same targets.\\
We have investigated the potential of the Pan-Planets survey to detect
VHN transiting M dwarfs brighter than i' = 18 mag. Assuming the
frequency of these objects is 5\%, we expect to find up to 3
detections in the first year and up to 9 detections observing the
same fields in the second year.\\
As a consistency check we modeled the OGLE-III Carina survey and found
2.18 VHJ, 1.46 HJ, 0.45 WJ, 0.12 VHS and zero VHN which is in
agreement with the 2 VHJ and 1 HJ and the zero WJ, VHS and VHN that
have been actually detected.
\begin{acknowledgements}
  We thank the referee Scott Gaudi for the constructive feedback. His
  comments and suggestions have helped us to identify the optimal
  survey strategy of the Pan-Planets project as well as improving the
  presentation.
\end{acknowledgements}
\end{document}